\documentclass[epsfig,graphics,twocolumn,showpacs,floatfix,mathbbm,prl,superscriptaddress]{revtex4}
\usepackage{amsmath,amssymb,graphicx,color}
\usepackage{float}
\newcommand{\PrYSO}{Pr$^{3+}$:Y$_2$Si{O$_5$}}
\newcommand{\YSO}{Y$_2$Si{O$_5$}}

\begin{document}

\title{Solid state source of non-classical photon pairs with embedded multimode quantum memory}
\pacs{03.67.Hk,42.50.Gy,42.50.Md}

\author{Kutlu Kutluer}
\author{Margherita Mazzera}
\email{margherita.mazzera@icfo.es}
\affiliation{ICFO-Institut de Ciencies Fotoniques, The Barcelona Institute of Science and Technology, Mediterranean Technology Park, 08860 Castelldefels (Barcelona), Spain}
\author{Hugues de Riedmatten}
\affiliation{ICFO-Institut de Ciencies Fotoniques, The Barcelona Institute of Science and Technology, Mediterranean Technology Park, 08860 Castelldefels (Barcelona), Spain}
\affiliation{ICREA-Instituci\'{o} Catalana de Recerca i Estudis Avan\c cats, 08015 Barcelona, Spain}

\date{\today}

\begin{abstract}
The generation and distribution of quantum correlations between photonic qubits is a key resource in quantum information science. For applications in quantum networks and quantum repeaters, it is required that these quantum correlations be stored in a quantum memory. In 2001, Duan, Lukin, Cirac, and Zoller (DLCZ) proposed a scheme combining a correlated photon-pair source and a quantum memory in atomic gases, which has enabled fast progress towards elementary quantum networks. In this letter, we demonstrate a solid state source of correlated photon pairs with embedded spin-wave quantum memory, using a rare-earth-ion doped crystal. We show strong quantum correlations between the photons, high enough for performing quantum communication. Unlike the original DLCZ proposal, our scheme is inherently multimode thanks to a built-in rephasing mechanism, allowing us to demonstrate storage of 11 temporal modes. These results represent an important step towards the realization of complex quantum networks architectures using solid-state resources.
\end{abstract}

\maketitle
Photonic quantum memories (QM) play a central role in quantum information science where they are used as quantum interfaces between flying and stationary qubits \cite{Afzelius2015}. 
Rare-earth ion doped solids (REIDS) are interesting candidates as QMs as they feature long coherence times \cite{Heinze2013,Zhong2015a} and prospects for integration \cite{Saglamyurek2011,Zhong2015,Marzban2015,Corrielli2016}. Most of the experiments to date in REIDS have demonstrated read-write quantum memories, where external quantum states of light are mapped on the QM \cite{Clausen2011,Saglamyurek2011,Rielander2014,Zhou2015,Seri2017}. This process requires the generation of single or entangled photons by an external source with demanding spectral properties to achieve strong interactions between the quantum light and the QM \cite{Rielander2016}. 
An alternative solution has been proposed \cite{Duan2001}, which combines a photon pair source and a quantum memory in a single physical system, an atomic gas. It is based on the creation of a single collective spin excitation (spin-wave) via off-resonant spontaneous Raman scattering, heralded by the detection of a Stokes photon. After a programmable storage time, the spin-wave can be transferred with high efficiency into a single anti-Stokes photon using a resonant read pulse. This scheme in principle leads to higher efficiencies than read-write QMs for the same optical depth, as the write-in stage is avoided \cite{Afzelius2015}. Since the first demonstration \cite{Kuzmich2003}, impressive progress has been realized, including the demonstration of elementary quantum networks \cite{Chaneliere2005,Eisaman2005,Chou2005, Chou2007, Choi2008, Yuan2008}, entanglement between four atomic ensembles \cite{Choi2010}, and long-lived quantum memories \cite{Radnaev2010,Yang2016}. These demonstrations have made cold atomic ensembles one of the most advanced systems to date for quantum networks applications. The DLCZ scheme has also been implemented with phonons in diamond \cite{Lee2012} and in a mechanical resonator \cite{Riedinger2016}.

Schemes to combine QMs and photon pair sources in rare-earth-ion doped solid-state ensembles have also been proposed \cite{Ledingham2010,Sekatski2011}. REIDS have much weaker optical transitions than atomic gases (their dipole moments is 2-3 orders of magnitude lower than alkali atoms), making off-resonant excitation challenging \cite{Goldschmidt2013}. A solution is to excite atoms on resonance to achieve stronger interaction. However, this leads to fast dephasing due to inhomogeneous broadening of the optical transition in the crystal, making a rephasing mechanism mandatory to recover the collective emission of the second photon \cite{Ledingham2010,Sekatski2011}. 
Early demonstrations of time-separated correlations between crystal and light field have been reported \cite{Ledingham2012,Beavan2012} using the scheme of Ref. \cite{Ledingham2010}, including entanglement between a light field and a solid-state spin-wave QM \cite{Ferguson2016}. In these experiments, quantum correlations were demonstrated in the continuous variable regime using heterodyne detection techniques. 

Here, we demonstrate a temporally multimode DLCZ-like scheme \cite{Sekatski2011} in the photon counting regime in a rare-earth-ion doped quantum memory. We generate pairs of correlated photons with a controllable delay and demonstrate quantum correlations between the photons for delays up to 20 $\mu $s. The photon pairs are created through spin-wave storage, effectively generating quantum correlations between single photons and single collective spin-waves in the crystal. We also show experimentally that combining the DLCZ scheme with rephasing techniques allows the creation of spin-waves into multiple temporal modes. The use of photon counting detection enabling discrete variable encoding, combined with the high quantum correlations demonstrated, makes our source of photon pairs with embedded quantum memory directly usable for quantum repeater schemes.

\begin{figure*}
	\centering{\includegraphics[width=1.9\columnwidth]{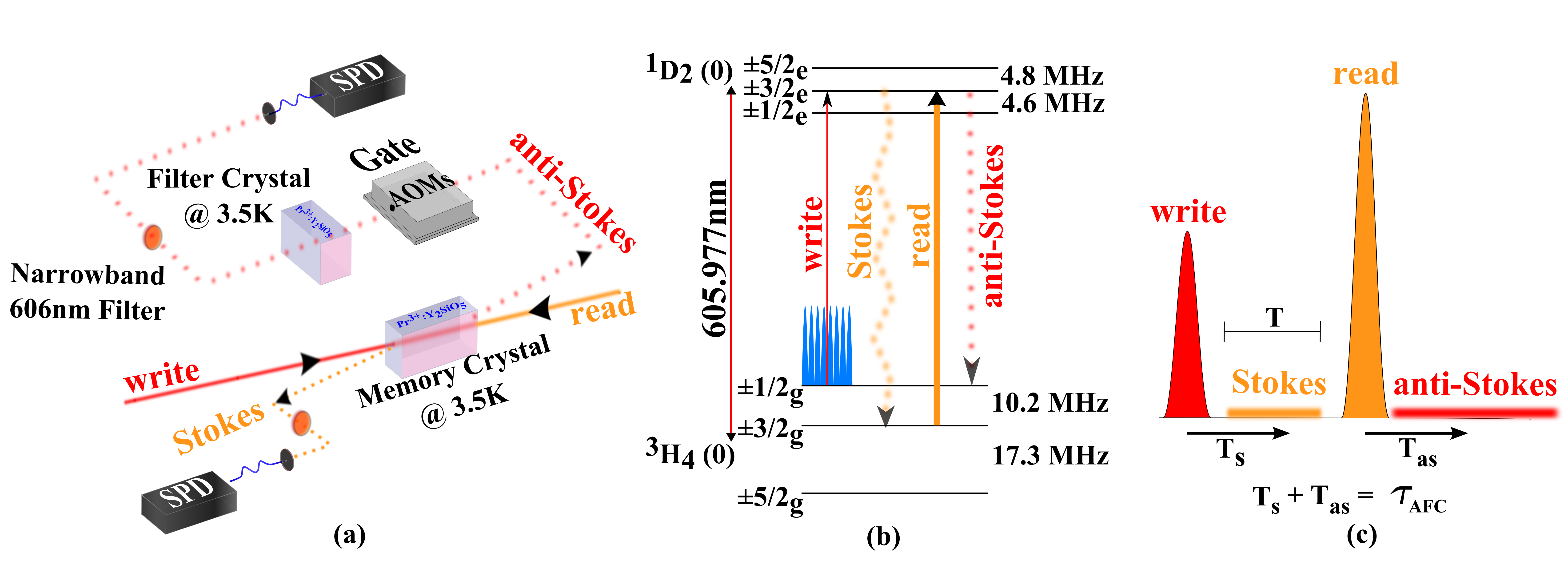}}
	\caption{(a) The experimental setup. The memory crystal, a 5 mm long, 0.05\% doped, Pr$^{3+}$:YSO, is hosted in a Montana Cryostation. The write and read pulses are counter-propagating. Their polarization is rotated along the D$_{2}$ crystal axis in order to maximize the absorption. Narrow-band filters at 600 nm (width 10 nm) are placed on both Stokes and anti-Stokes modes. The Stokes photons are fiber coupled to the detector with about 75 $\%$ transmission. The anti-Stokes photons are first temporally gated by two acoustic optical modulators (AOMs) and later spectrally filtered by a 1 MHz-wide spectral hole at the 1/2$ _{g} $ - 3/2$ _{e} $ transition frequency burnt in a second Pr$^{3+}$:YSO crystal, 3-mm long, also at 3.5 K. The total transmission in the anti-Stokes arm, from the cryostat to the detector, is typically $24\,\%$. We use two Silicon single photon detectors (SPD) for the detection of both photons.  (b) Hyperfine splitting of the first sub-levels (0) of the ground $^{3}$H$_{4}$ and the excited $^{1}$D$_{2}$ manifolds of Pr$^{3+}$ in \YSO. The AFC structure (blue comb) is prepared with the read mode. (c) Temporal pulse sequence. The write pulses are 1 $\mu $s long (FWHM), and have a negative frequency chirp with a hyperbolic tangent waveform of 800 kHz. The read pulses, sent 8 $\mu $s after the write pulses, have power of 24 mW, duration of 900 ns (FWHM), and frequency chirp of +800 kHz with a hyperbolic tangent waveform. $T_s$ ($T_{as}$) is the time separation between a Stokes (anti-Stokes) photon and the write (read) pulse. T is the width of the Stokes detection window. For one AFC preparation we send 500 write-read pairs every 313 $ \mu$s for $P_w \leq 128 \mu $W. For higher $P_w$ we decrease the number of trials to prevent deterioration of the AFC. } 
\label{setup}
\end{figure*}

\begin{figure*}
\centerline{\includegraphics[width=1.9\columnwidth]{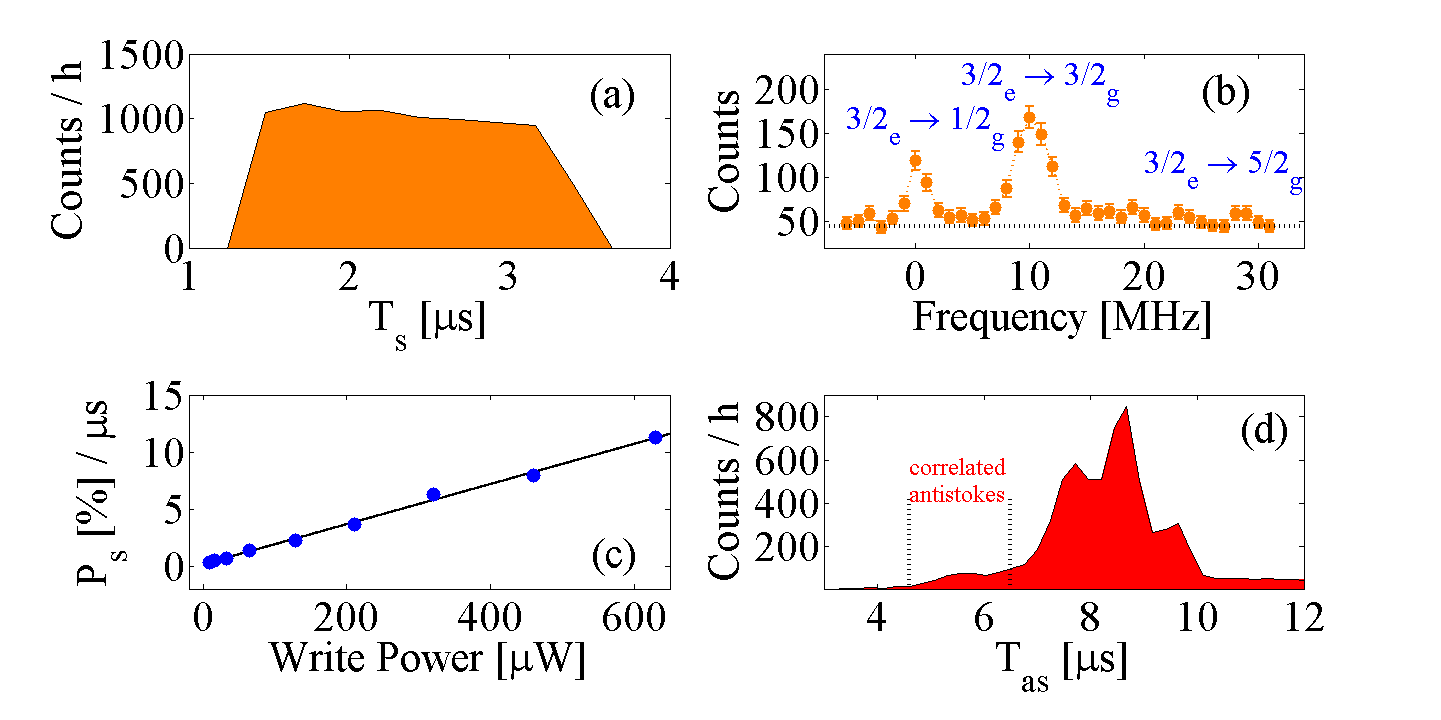}}
\caption{(a) Stokes count rate in a 2 $ \mu $s detection window, for $P_w$ = 16 $ \mu $W. (b) Stokes counts as a function of the position of a 1 MHz-wide spectral hole burnt in the filter crystal. The 0 frequency corresponds to the frequency of the write pulse. The dotted gray horizontal line is the noise given by the detector dark counts. (c) Probability to create Stokes photons as a function of the write pulse power. The solid curve is a linear fit of the experimental data points. (d) Anti-Stokes count rate. The dotted vertical bars indicate the temporal mode where the anti-Stokes photons correlated to the Stokes photons of panel (a) should lie to satisfy the condition T$_s$+T$_{as}$ = $\tau_{AFC}$. This histogram shows a peak at around 8.5 $ \mu $s originated from the second echo of the write pulse leaked into the anti-Stokes mode. Note that this peak is not in the temporal mode of the correlated anti-Stokes photons. Nevertheless, from the Gaussian fit of the two peaks we infer a contribution to the anti-Stokes counts of around $4\%$. }

\label{Fig2}
\end{figure*}

In our experiment, the rephasing mechanism uses the Atomic Frequency Comb (AFC) scheme. It relies on the spectral tailoring of the inhomogeneous absorption profile as comb like structures with a spectral periodicity $\Delta$ which rephase the ions at time $\tau_{AFC}=1/\Delta $ and lead to a collective emission \cite{Afzelius2009,Riedmatten2008a}.
The sample used is a \PrYSO\, (Pr$^{3+}$:YSO) crystal cooled down to 3.5 K. Fig. \ref{setup} shows the experimental setup (a) and the relevant energy level scheme of the crystal (b). We first prepare an AFC on the 1/2$ _{g} $ - 3/2$ _{e} $ transition, with $\tau_{AFC} $ = 8 $ \mu $s. At the same time the 3/2$ _{g} $ state is emptied for the single spin excitation (see Appendix). Then we start a progression of trials, each consisting of the pulse sequence depicted in Fig. \ref{setup}(c). We send write pulses resonant with the AFC and, after 1.4 $ \mu $s, we open the gate to detect the Stokes photons. These are emitted in the whole solid angle, but we only collect the Stokes field backward at an angle of $\sim 4\,\mathrm{^{\circ}}$ with respect to the write mode. With our filtering system we mostly collect only photons resonant to the $^{1}$D$_{2}(0) \rightarrow ^{3}$H$_{4}(0)$ transition. Then we unconditionally send strong read pulses. The write and read pulses are counter-propagating, therefore, due to the phase matching condition ($ \overrightarrow{k} _{as} = \overrightarrow{k}_{w} + \overrightarrow{k}_{r} - \overrightarrow{k}_{s}$) and collective interference, the anti-Stokes photons are emitted counter-propagating with respect to the detected Stokes photon. The anti-Stokes photons are temporally and spectrally filtered before being steered to the detection stage. 
 
We start our measurements by characterizing the light emitted in the Stokes mode. Fig. \ref{Fig2}(a) shows the temporal Stokes histogram in a 2 $ \mu $s window, for a write power $P_w$ = 16 $ \mu $W. 
Fig. \ref{Fig2}(b) shows the spectrum of the emitted Stokes field from which we infer that the hyperfine branching ratio of the Stokes photons, $ \beta_{BR} $, at the relevant 3/2$ _{g} $ - 3/2$ _{e} $ transition frequency is about 60\% (see Appendix). We also measure the probability to generate a Stokes photon $P_s$ as a function of $P_w$, as shown in Fig. \ref{Fig2}(c). We observe a linear dependence, as predicted in \cite{Sekatski2011} for $P_s \ll 1$. These observations suggest that the light emitted in the Stokes mode comes from the direct excitation of the ions by the write pulse. Fig. \ref{Fig2}(d) also shows the time histogram of the anti-Stokes mode, after sending the read pulse. 

\begin{figure*}
\centering{\includegraphics[width=1.9\columnwidth]{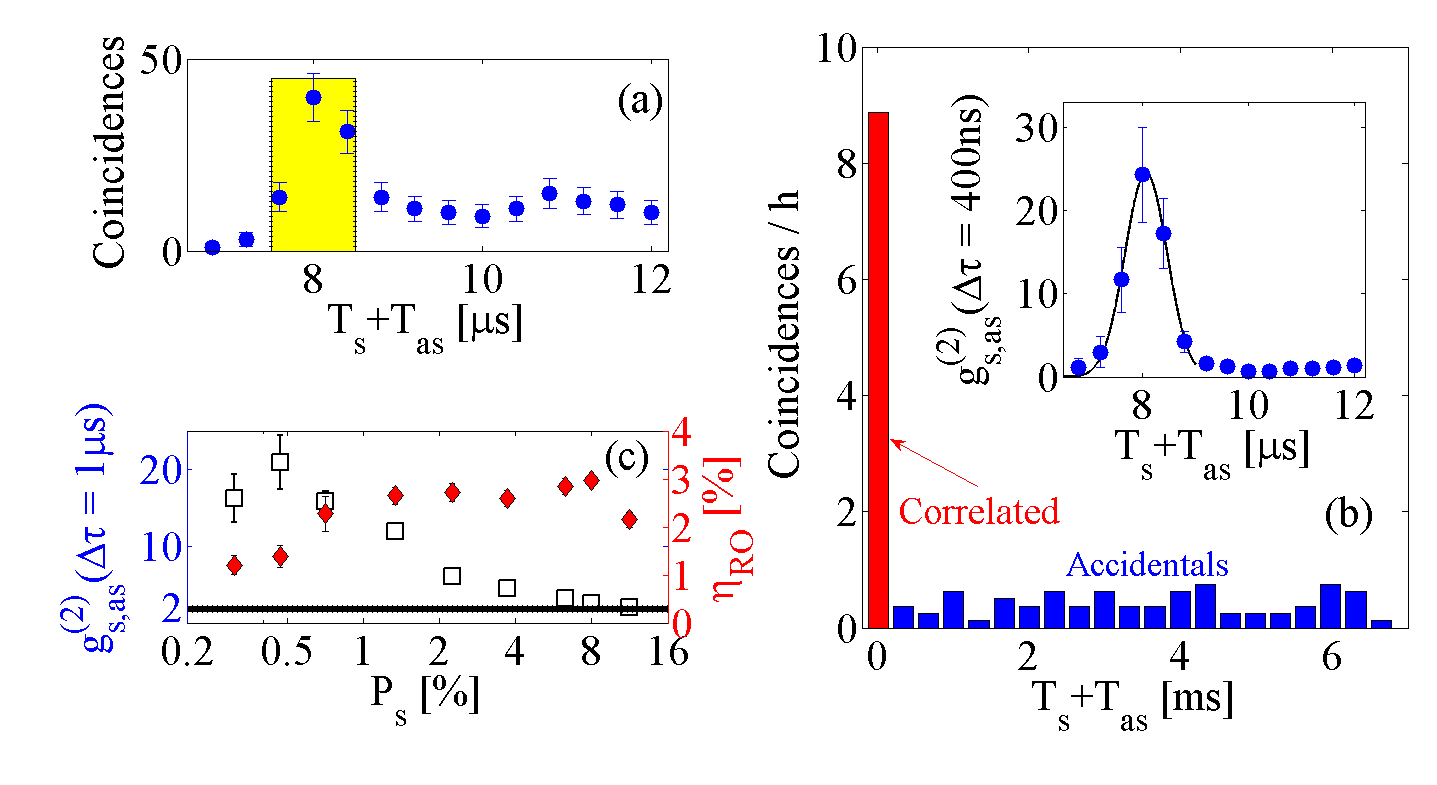}}
\caption{(a) Coincidence counts between Stokes and anti-Stokes photons as a function of the sum $T_s$+$T_{as}$. The bin-size is $\Delta \tau$ = 400 ns. The shaded area indicates the time window used to calculate the $g^{(2)} _{s,as}(\Delta \tau = 1 \mu$s) value. We also observe a smaller peak at 11 $\mu $s, due to the noise generated by the second AFC echo of the write pulse (see Fig. \ref{Fig2}(d)). However, this peak is also present when Stokes and anti-Stokes photons are detected in different trials and therefore does not correspond to correlated photons. 
(b) Coincidence counts per hour between Stokes and anti-Stokes photons in the 1 $\mu$s wide temporal mode around T$_s$+T$_{as}$ = $\tau_{AFC}$ = 8 $\mu$s in the same storage trial (bar at 0) and in different storage trials separated by multiples of 313 $\mu$s. The $g^{(2)} _{s,as}$ value is calculated as the ratio between the coincidences in the same storage trial and the average of the coincidences in different storage trials. The inset shows the peak in the $g^{(2)} _{s,as}$ with bin-size of $\Delta \tau$ = 400 ns. The fit of the correlation peak to a Gaussian curve, done over a 2 $\mu$s window around the peak, is also shown. (c) $g^{(2)} _{s,as}$ value (squares) and read-out efficiency (diamonds) as a function of the Stokes creation probability, P$_s$, calculated by the raw detection counts back propagated at the crystal using known losses. The black horizontal line sets the classical threshold for the $g^{(2)} _{s,as}$ given by the Cauchy-Schwarz inequality. }
\label{Fig3}
\end{figure*}

We now look for coincidence detection between the Stokes and anti-Stokes modes. The Stokes photon which heralds a single-spin excitation is emitted from the ions which have spent in the excited state a time $T_s$. After a storage time $t_S$ in the spin state, the read pulse then transfers the collective excitation from the spin-state back to the excited state. At that moment, the phase evolution in the excited state resumes, and the correlated anti-Stokes photon will be emitted at time $T_{as}$ after the read pulse, such that $T_s+T_{as}=\tau_{AFC}$. We record the detection times of the Stokes ($T_s$) and anti-Stokes ($T_{as}$) photons, and plot a histogram of the coincidences as a function of the sum $T_s+T_{as}$. For correlated pairs, we therefore expect a coincidence peak at $T_s+T_{as}=\tau_{AFC}$.  Fig \ref{Fig3}(a) shows such a histogram for $P_w$ = 16 $\mu W$, where we see a clear correlation peak around $\tau_{AFC}$=8 $\mu s$. 

To quantify the correlation between Stokes and anti-Stokes photons, we measure intensity correlation functions. The second order cross-correlation function $g^{(2)} _{s,as}$ is defined as $g^{(2)} _{s,as} = p_{s,as}/(p_{s} \cdot p_{as})$, where  $p_{s, as}$ is the probability to detect a coincidence between Stokes and anti-Stokes photons and $p_{s}$ ($p_{as}$) is the probability to detect a Stokes (an anti-Stokes) photon. To infer $g^{(2)} _{s,as}$, we measure the number of coincidences in a time window $\Delta \tau$ around $T_s+T_{as}=\tau_{AFC}$ in the same trial, and we compare this number with the accidental coincidences recorded for Stokes and anti-Stokes photon emitted in different independent trials. An example is shown in Fig. \ref{Fig3}(b) for $P_w$= 16 $\mu W$ and $\Delta \tau = 1 \mu$s, which includes around 76 $\%$ of the total peak counts. For this particular example, with average t$_S = 5.6 \mu $s, we find $g^{(2)} _{s,as} (\Delta \tau = 1 \mu s)$ = 21 $\pm$ 4. This is much higher than the threshold of about 6 enabling the violation of a Bell inequality if at least two modes are stored \cite{Riedmatten2006}. The inset in Fig. \ref{Fig3}(b) shows a zoom on the correlation peak for a smaller $\Delta \tau$ = 400 ns, leading to $g^{(2)} _{s,as} (\Delta \tau = 400 $ns) = 24 $\pm$ 6. The shape of the normalized correlation peak is fitted with a Gaussian curve and a temporal FWHM of (940 $\pm$ 100) ns is extracted. This is close to the expected width corresponding to the write pulse duration convoluted with the time-bin size. 

To further characterize our system, we measure  $g^{(2)} _{s,as}(\Delta t=1 \mu s)$ as a function of $P_s$ (Fig. \ref{Fig3}(c)), which can be adjusted by tuning $P_w$ (see Fig. \ref{Fig2}(c)). We observe a reduction of the $g^{(2)} _{s,as}$ value when $P_s$ is increased, as expected for a DLCZ-like process. At lower $P_s$, the rate of Stokes photons becomes comparable to the noise and the value of $g^{(2)} _{s,as}$ decreases. We also measure the read-out efficiency $\eta_{RO}=(p_{s,as}-p_{s,as}^{acc})/p_s$ as a function of $P_s$, where $p_{s,as}^{acc}$ is the accidental coincidence probability. We observe a linear increase up to $P_s$ = 1 $\%$, due to the noise in the Stokes mode. Afterwards it stays constant around 3\%, close to the expected value (see Supplemental Materials). This value is comparable to the read-out efficiency achieved in the latest demonstration of continuous variable entanglement between light and spin waves implemented in the same system \cite{Ferguson2016}.

To prove unambiguously the non-classical correlations between the two photons, we use the Cauchy-Schwarz inequality: 
\begin{center}
	$R = \frac{(g^{(2)} _{s,as})^{2}}{g^{(2)} _{s,s} \cdot g^{(2)} _{as,as}} \leq 1$, 
\end{center}

where $g^{(2)} _{s,s}$ and $g^{(2)} _{as,as}$ are second order auto-correlation functions of Stokes and anti-Stokes photons, respectively. To measure these quantities, we introduce a 50/50 fiber splitter and two SPDs in the Stokes (anti-Stokes) arm and proceed in a similar way as for the cross-correlation. 
With $P_w$ = 64 $ \mu W $ we find $g^{(2)} _{s,s}(\Delta t = 1 \mu s) = 1.85 \pm 0.36$, $g^{(2)} _{as,as}(\Delta t = 1 \mu s) = 1.75 \pm 0.57$, and $g^{(2)} _{s,as}(\Delta t = 1\mu s) = 11.9\pm 0.9$. Eventually we find $ R = 44 \pm 20$, which exceeds the classical limit by more than 2 standard deviations. 

\begin{figure}
	\centering{\includegraphics[width=1\columnwidth]{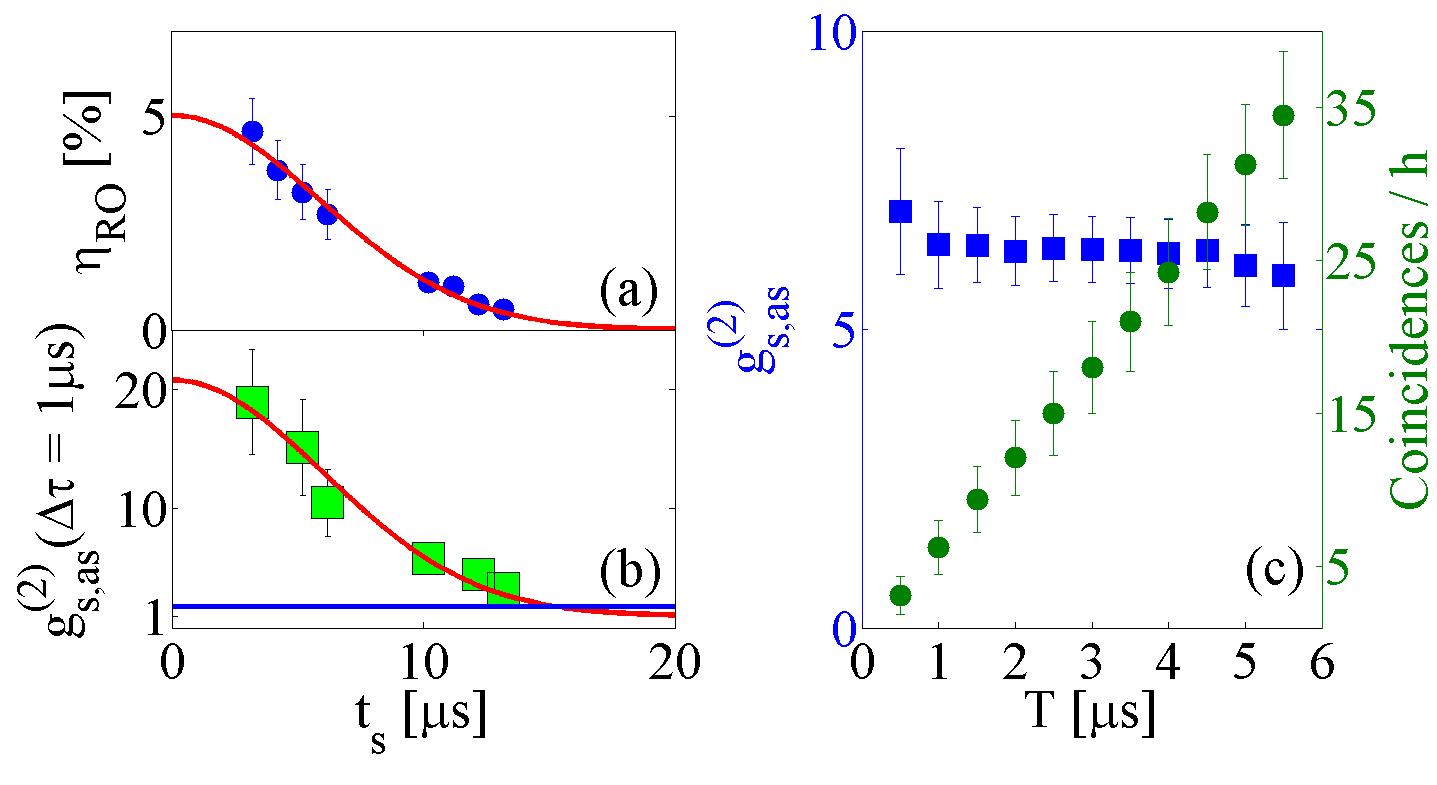}}
	\caption{Read-out efficiency (a) and $g^{(2)} _{s,as}(\Delta t = 1 \mu s)$ cross-correlation values (b) as a function of the storage time in the spin state t$_S$. The red line is the fit of the experimental data to a Gaussian decay which gives a 1/e decay time of (8.3 $\pm$ 0.8) $\mu $s, corresponding to a spin inhomogeneous linewidth of 45 $ \pm 2 $ kHz (44  $ \pm 4 $ kHz for $g^{(2)} _{s,as}$). For this curve, we do not subtract the accidental counts. The blue horizontal line in (b) sets the classical threshold for the $g^{(2)} _{s,as}$ given by the Cauchy-Schwarz inequality.
(c) $g^{(2)} _{s,as}(\Delta t = 1 \mu s)$ cross-correlation (squares) and coincidence counts per hour (circles) between the Stokes and anti-Stokes photons as a function of Stokes window size, T. The total Stokes window size is 5.5 $\mu$s. In the data processing stage we adjust the stokes window size as multiples of the duration of the write pulse (FWHM = 500 ns). The values are averages of all possible windows for different window sizes, e.g. 11 for a 500-ns window, 9 for a 1500-ns window, etc. In this measurement P$_w$ = 64 $\mu$W. The read pulse is sent 15 $\mu$s after the write pulse. The decrease on $g^{(2)} _{s,as}$ with respect to the one in Fig. \ref{Fig3}(a) is due to the longer spin-state storage time and less efficient read-out for shorter write pulse ($\eta_{RO}$ = 1 $\%$).}
	\label{Fig4}
\end{figure}

To show that the delay between the Stokes and anti-Stokes photons can be controlled, we measure the retrieval efficiency as a function of the storage time in the spin-state $t_S$, as shown in Fig. \ref{Fig4}(a). The data are fitted with a Gaussian curve, from which we extract a 1/e decay constant of (8.3 $\pm$ 0.8) $\mu $s. 
This decay is likely due to inhomogeneous broadening of the spin transition. The same measurement in a standard spin-wave AFC experiment in this crystal gives a similar value (9.9 $\pm$ 1.5) $\mu$s. The decay in $\eta_{RO}$ also affects $g^{(2)} _{s,as}$, as shown in Fig \ref{Fig4}(b). Nevertheless, we observe non-classical correlation for $t_S$ up to 12 $\mu $s, corresponding to a total storage time of $t_S+\tau_{AFC}$ = 20 $\mu $s. 

Finally, we discuss the temporal multimode nature of our source. The detection of a Stokes photon at a different time $T_s'$ creates a spin-wave that will also rephase at a different time $T_{as}'$, still preserving $T_s'+T_{as}'=\tau_{AFC}$. The maximal number of modes is given by $N_{m}=T/\Delta \tau$, where T is the detection window of the Stokes photons. Note that there is a trade-off between the number of modes and the read-out efficiency of each mode. The efficiency is maximized if $\Delta \tau$ is bigger than the correlation peak. In our case, we choose $\Delta \tau$ comparable to the FWHM of the peak. Fig. \ref{Fig4}(c) shows the coincidence count rate and the g$^{(2)} _{s,as}$ as a function of $T$. 
We observe a linear increase in coincidence count rate while increasing $T$. At the same time, the $g^{(2)} _{s,as}$ values stay constant. This shows an important feature of our scheme: adding more temporal modes increases the coincidence rate but does not decrease the correlation. The maximal number of modes in our current experiment is $N_{m}=T/\Delta \tau$ = 11 modes, a factor of 5 improvement with respect to previous multimode DLCZ-like experiments \cite{Albrecht2015,Ferguson2016}. 

Several improvements to our system are possible. The Stokes probability could be improved, without resorting to higher write pulse powers. For example longer crystals or, to some extent, higher concentrations could be employed to achieve higher optical depths. The light-atom interaction could also be enhanced with external cavities. These measures would also be beneficial for the read-out efficiency. The latter could be further boosted by filtering the Stokes photon for suppression of wrong heralds, by improving the quality of the AFC for better rephasing, and by optimizing the transfer efficiency.
The storage time could be greatly improved using spin-echo techniques \cite{Jobez2015}, with prospects of up to tens of seconds in our crystal \cite{Heinze2013} or hours in Europium doped crystals \cite{Zhong2015a}.  Even though the emitted photons are at 606 nm, it is possible to efficiently convert them to telecom C-band wavelength using quantum frequency conversion techniques \cite{Maring2014}. 

Our observation of non-classical correlations between photon pairs emitted from a solid-state ensemble with a controllable delay by using photon counting techniques represents an important achievement towards scalable quantum repeater architectures. The demonstrated inherent temporal multimodality of our protocol also paves the way towards heralded entanglement of remote solids with temporal multiplexing, which can greatly enhance the distribution rate of entanglement \cite{Simon2007}.

\textit{Note added} - We note that a similar work demonstrating the AFC- DLCZ protocol in an europium doped crystal has been performed independently and is published alongside the present Letter \cite{Laplane2017}.

\textbf{Acknowledgements.} We acknowledge financial support by the ERC starting grant QuLIMA, by the Spanish Ministry of Economy and Competitiveness (MINECO) and Fondo Europeo de Desarrollo Regional (FEDER) (FIS2015-69535-R), by MINECO Severo Ochoa through grant SEV-2015-0522, by AGAUR via 2014 SGR 1554, by Fundaci\'o Cellex, and by CERCA Programme/Generalitat de Catalunya.

\hrulefill
\newpage


\appendix

\section{Appendix}

\maketitle
In the present Supplemental Material we provide detailed information about the preparation of the Atomic Frequency Comb (AFC) and the calculation of the read-out efficiency.

\section{AFC Preparation}

The method we use for tailoring the absorption profile of the Pr$^{3+}$:YSO crystal as an AFC structure consists of two main steps. The first step is the creation of a single class 4 MHz absorptive feature. To do this we start by creating a 18 MHz-wide transparency window by sweeping a 24 mW pulse within the inhomogeneous profile on the crystal. This sweep has the effect of emptying the 1/2$ _{g} $ and 3/2$ _{g} $ states of a given class of atoms. A burn-back process then creates a 4 MHz-wide absorption feature at the frequency of the 1/2$ _{g} $ - 3/2$ _{e} $ transition but also populates the 3/2$ _{g} $ state, which will be used for the single spin excitation. Thus, we clean it by sending strong pulses resonant with the 3/2$ _{g} $ - 3/2$ _{e} $ transition. This operation also contributes to remove any absorption features associated to other atomic classes. 

The second step is the shaping the AFC structure on the single class feature. As discussed in \cite{Bonarota2010}, the optimum AFC peak shape to maximize the efficiency is square. We prepare a waveform of a square AFC structure by considering the OD of crystal and the excited state storage time. The Fourier transform of the desired AFC gives the temporal waveform of the pulse train that we use to shape the single class absorption feature \cite{Jobez2016}. While we burn the AFC at the 1/2$ _{g} $ - 3/2$ _{e} $ transition frequency, some ions are also coherently driven back to the 3/2$ _{g} $ state. Thus, we proceed by sending cleaning pulses at the frequency of the 3/2$ _{g} $ - 3/2$ _{e} $ transition. The AFC shaping and spin-state cleaning procedures are repeated 1100 times. All the preparation pulses are sent along the read spatial mode.

\section{Read-out Efficiency}

The measured value of the read-out efficiency can be compared to the expected one with our experimental parameters, which can be estimated as $\eta_{RO}^{exp} = \eta_{RP} \cdot \eta_{reph} \cdot \eta_{decoh} \cdot \beta_{BR} \cdot \beta_{G}$, where $ \eta_{RP} $ is the transfer efficiency of the read pulse, $ \eta_{reph} $ is the rephasing efficiency of the AFC, $ \eta_{decoh} $ is the decoherence in the spin-state, $ \beta_{BR} $ is the branching ratio, and $ \beta_{G} $ is the fraction of the pulse counts in the detection window $\Delta \tau$. We calculate $\eta_{RO}^{exp}$ by measuring $\eta_{RP}, \eta_{reph}, \eta_{decoh}$, and $\beta_{BR}$ separately. 
In order to find the $\eta_{RP}$ we conduct \textit{classical} DLCZ-AFC experiment. We prepare the AFC and send a write pulse but, instead of Stokes photons detection, a weak probe pulse is sent which transfers a small part of the ions from the excited state to the spin state. The weak probe is sent from the anti-Stokes spatial mode in the temporal window of the Stokes detection. The stimulated transfer generates photon emission (gain) during the weak pulse which is proportional to the number of ions transferred to the spin state. The strong read pulse transfers the population back to the excited state and the collective emission occurs. By detecting the gain of the weak probe and the collective emission of the AFC we extract $\eta_{RP} = 40\%$.

We calculate the decoherence in the spin state by taking into account the inhomogeneous spin broadening measured in the Fig.4(a). For a 2 $\mu$s-wide Stokes window, the average spin state storage time is 5.6 $\mu$s which causes a 64\% decoherence. 

The AFC efficiency can be defined as $\eta_{AFC} = \eta_{write} \cdot \eta_{rephasing} \cdot \eta_{loss}$, where we measure $\eta_{AFC} = 17\%$. The efficiency of the write pulse is defined as $\eta_{write} \approx 1 - e^{-d/F}$ where d is the absorption depth and F is the finesse of the AFC. The loss in the AFC protocol is defined as $\eta_{loss} = e^{-d_{0}}$ where $d_{0}$ is the background absorption. We measure an average d of 5.4, F of 4.4 and $d_{0}$ = 0.4 and finally find $\eta_{rephasing} = 36\%$.
 
We characterize the Stokes photons frequency by spectral filtering with the help of the filter crystal that we use in the experiment. We send the write pulse from the read mode therefore we detect the Stokes photons after they pass through the filter crystal. We shift the 1 MHz hole burnt in the filter crystal at different frequencies. The Stokes photons detected at the 3/2$ _{g} $ - 3/2$ _{e} $ transition frequency are the 60\% of the overall detections. 

When measuring the read-out efficiency, we consider $\Delta \tau = 1 \mu$s, which includes a fraction $ \beta_{G} $ = 76\% of the coincidence peak. 

Finally the expected readout efficiency is $\eta_{RO}^{exp} = 40\% \cdot 64\% \cdot 36\% \cdot 60\% \cdot 76\% = 4.2\%$.

\end{document}